\documentclass{article}

\usepackage{times}
\usepackage{epsfig}
\usepackage{graphicx}
\usepackage{amsmath}
\usepackage{amssymb}
\usepackage{booktabs}
\usepackage{cite}
\usepackage[ruled,vlined,linesnumbered]{algorithm2e}
\usepackage{caption}
\usepackage{subcaption}
\usepackage{adjustbox}
\usepackage{arydshln}
\usepackage{multirow}

\usepackage{ marvosym }

\def\etal{\textit{et al.,}~}
\def\ie{\textit{i.e.,}~}
\def\eg{\textit{e.g.,}~}

\usepackage[breaklinks=true,bookmarks=false]{hyperref}

\begin{document}

\title{Learning to Jointly Deblur, Demosaick and Denoise Raw Images}

\renewcommand{\thefootnote}{\fnsymbol{footnote}}

\author{Thomas Eboli$^{\ast,\dagger}$
\and
Jian Sun$^{\ddagger}$
\and
Jean Ponce$^{\ast,\dagger}$
}

\maketitle

\begin{abstract}
    We address the problem of non-blind deblurring and demosaicking of noisy 
    raw images.
    We adapt an existing learning-based approach to RGB image deblurring to
    handle raw images by introducing a new interpretable
    module that jointly demosaicks and deblurs them.
    We train this model on RGB images converted into raw ones following
    a realistic invertible camera pipeline.
    We demonstrate the effectiveness of this model over two-stage approaches
    stacking demosaicking and deblurring modules on quantitive benchmarks.
    We also apply our approach to remove a camera's inherent blur
    (its color-dependent point-spread function)
    from real images, in essence deblurring sharp images.
\end{abstract}

\section{Introduction}
\footnotetext [1]{INRIA.}
\footnotetext [2]{D\'epartement d'informatique de l'ENS, Ecole normale
sup\'erieure, CNRS, PSL Research University, France.}
\footnotetext [3]{Department of Information Science, School of Mathematics and Statistics, Xi'an Jiaotong University, P.R.China.}
\footnotetext{\Letter: Correspondance at \texttt{thomas.eboli@inria.fr}}

\renewcommand{\thefootnote}{\arabic{footnote}}

The goal of this work is to deblur, denoise and demosaick raw
images.
Raw data is important since it captures the most direct information
we have about the observed scene, before any digital post-processing
such as color transformations and gamma correction \cite{brooks19unprocessing}.
An important application
is the removal of the optical aberrations introduced by the 
lens point-spread function (PSF).
Indeed, any photograph, even perfectly focused and in the absence of any motion,
contains some blur caused by its optics, ranging from geometric
distortions to chromatic aberrations \cite{schuler11correction, yue15blind}.
Removing these artifacts is a (little explored) instance of joint image demosaicking
and non-blind deblurring addressed in this presentation.

Most approaches to image deblurring focus on sophisticated priors \cite{krishnan09fast, zoran11from},
convolutional neural networks (CNNs) \cite{nah17deep, tao18scale},
or a combination of both \cite{chen17trainable, zhang20deep, eboli20end2end, kruse17learning}.
Recent algorithms are robust to various noise levels
\cite{kruse17learning},
large \cite{eboli20end2end, zhang20deep} and even approximate kernels \cite{pan18dark}, but
they often ignore several stages
of the camera pipeline connecting the analog image in the focal plane to the digital blurry image recorded by the camera.

Blur is caused by various, color-dependent optical phenomena
\cite{schuler11correction, yue15blind}, camera and/or scene motion, and 
spatial, spectral and temporal integration over the pixel area. 
In particular, a single grey value is typically recorded at each pixel
according to the Bayer pattern to form the final {\em raw} image.

Raw images are interpolated with filtering techniques
\cite{li08survey, menon07demosaicing} or learning-based approaches \cite{gharbi16deep, kokkinos19iterative}
into linear RGB (aka linRGB) images \cite{plotz17benchmarking, brooks19unprocessing}. 
This {\em demosaicking} operation is often highly non-linear.
Sensor noise follows a statistical
model whose parameters are estimated empirically from raw images \cite{foi08practical} or 
learnt with a neural network on a corpus of image pairs
\cite{abdelhamed19noise}, which is much 
more realistic than a Gaussian model.
LinRGB images are finally converted into the standard RGB (aka sRGB)
format through an image signal processing (ISP) pipeline 
\cite{brooks19unprocessing}.

\begin{figure*}
    \centering
    \includegraphics[scale=0.58]{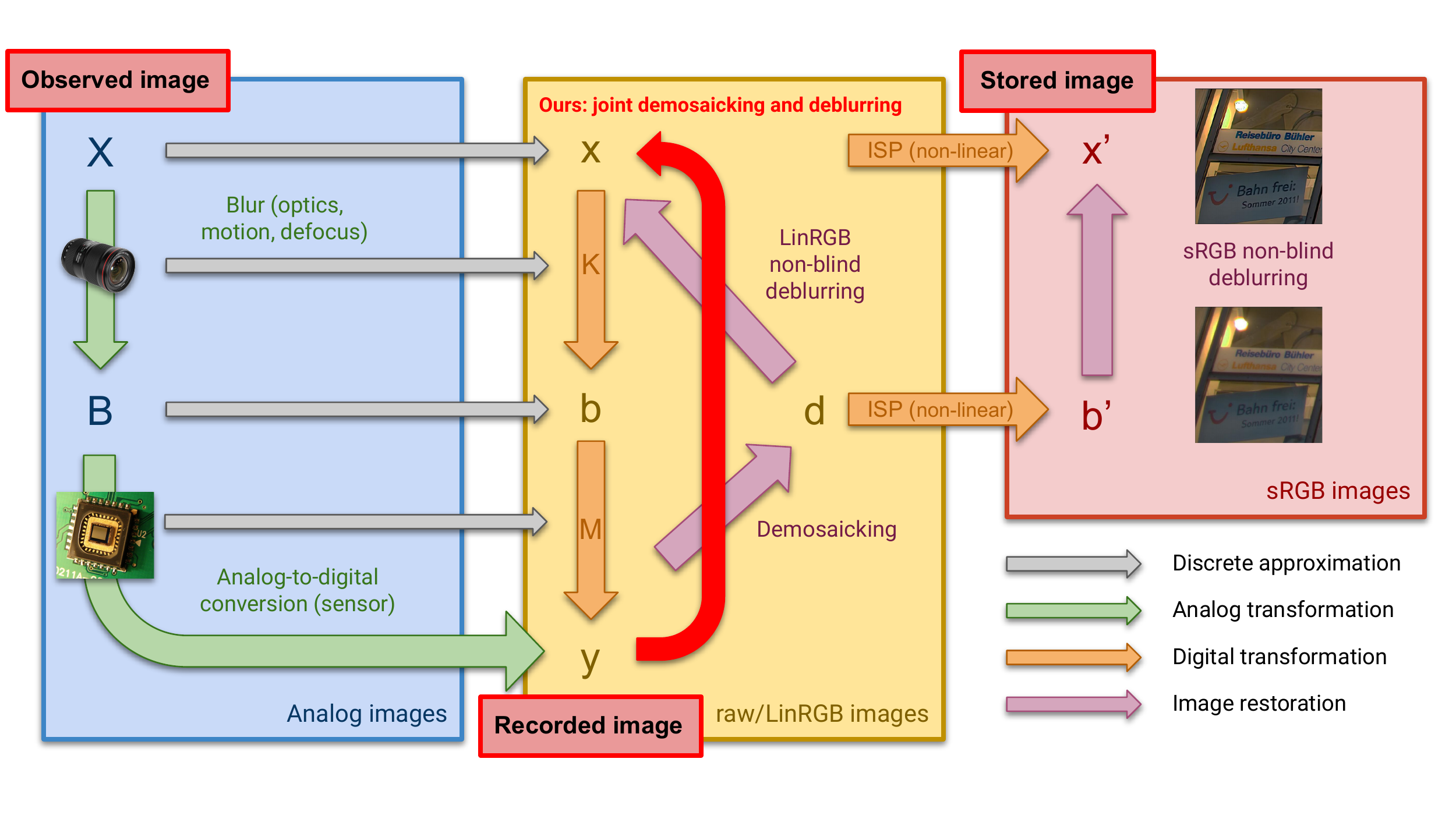}
    \caption{Starting from an observed image $X$ in its focal plane, a 
    digital camera records a blurry, mosaicked and noisy raw image $y$.
    Several processing stages are required to display a sharp sRGB image $x^\prime$. We predict from $y$ a sharp linRGB image 
    $x$, which consists in inverting the linear operator $MK$.
    We do not need an intermediate demosaicked image $d$ that 
    might contain prediction errors, Moir\'e artifacts and smoothed 
    details.
    The image $x$ is further converted into sRGB format with an ISP pipeline.}
    \label{fig:forward_model}
\end{figure*}

This complex process suggests that the classical model of convolving a sharp image
with a linear filter to form its blurry version can be improved
to better fit real digital cameras.
We thus start from a raw, blurry and noisy image to
predict a sharp and denoised linRGB image.
We use a realistic image formation model, 
embed it into a penalized energy term, and
unroll a few stages of an iterative solver within a parametric function
inspired by \cite{zhang20deep} and trained with samples preprocessed
with a modified variant of the linRGB-to-sRGB conversion 
pipeline from \cite{brooks19unprocessing}.
We finally apply this model to optical aberrations removal
from images taken
with high-end cameras whose PSF has been estimated separately.

Our main contributions can be summarized as follows:
\begin{itemize}
    \item We introduce a joint deblurring, demosaicking and denoisng 
    formulation motivated by a realistic camera pipeline;
    \item a penalized energy based on this forward model optimized with
     a splitting method. We unroll a few stages of this iterative solver 
     within a parametric function
     trained on blurry, mosaicked and noisy images generated with \cite{brooks19unprocessing};
    \item we present an experimental comparison with two-stage methods demonstrating
    the benefits of our approach and;
    \item we present an application of the proposed model to the removal
    of blur and chromatic aberrations caused by a camera's PSF 
    on both synthetic and real images.
\end{itemize}

\section{Related work}

\paragraph{ISP pipeline models.}
Previous approaches for modelling realistic image noise
have focused on estimating an empirical noise distribution
in linRGB and raw images.
Foi \etal \cite{foi08practical} model noise as a Poissonian-Gaussian 
mixture model, also used in \cite{plotz17benchmarking, abdelhamed18high} for
generating realistic images. 
Since noise is better modelled on raw/linRGB images, several
approaches for image denoising directly work on mosaicked raw images,
yielding joint demosaicking and denoising methods 
\cite{khashabi14joint, klatzer16learning, gharbi16deep, kokkinos19iterative}.
In particular, \cite{khashabi14joint, klatzer16learning} predict denoised and
demosaicked linRGB images but compute the training error on its sRGB versions
since images are ultimately rendered in this color format.
Such supervision for CNNs demands large corpuses of aligned image pairs that are
hard to obtain in general for image restoration tasks. 
Brooks \etal \cite{brooks19unprocessing} simulate a forward model approximating
a general ISP pipeline
for generating raw/linRGB degraded images from sRGB clean ones. It is made invertible
such that supervision of CNNs predicting linRGB/raw images with sRGB targets is possible.
Likewise, Nah \etal \cite{nah17deep} invert gamma correction to 
build blurry training data but most deblurring approaches
only tackle sRGB images.

\paragraph{Image deblurring.}
Classical approaches to deconvolution (aka non-blind deblurring) include
Wiener filtering\cite{gonzalez92image} and
variational methods \cite{krishnan09fast, zoran11from} built with handcrafted or learnt priors on image gradients and patches.
Optimization is traditionally carried out with a technique such as
half-quadratic splitting (HQS) \cite{geman92constrained}.
HQS has become the backbone of parametric functions
unrolling a few iterations to build interpretable methods with learnable priors
\cite{schmidt13discriminative, kruse17learning, zhang20deep, eboli20end2end}.
Another trend \cite{schuler13machine, dong20deep} is to learn a CNN that compensates
the ringing artifacts of Wiener filtering.
Finally, some methods \cite{nah17deep, tao18scale} give up the need for an image
forward model by training highly-engineered architectures with large 
collections of blurry and sharp images as supervision for blind image deblurring.

\paragraph{Joint deblurring and demosaicking.}
A classical approach to this problem is to combine demosaicking and a non-blind deblurring approaches
into a two-stage model \cite{schuler11correction}.
Joint demosaicking and deblurring can also be modelled as solving an 
inverse problem with edge-preserving priors and 
spectral regularization \cite{soulez09joint, trimeche05multichannel}.
Liang \etal \cite{liang20raw} synthesize training pairs composed of blurry raw 
and sharp sRGB images from videos taken with a GoPro camera
and train a multi-branch CNN to address joint blind deblurring
and demosaicking.
Such data is hard to collect and lacks diversity because of the 
small number of different scenes recorded.
We instead use the ISP pipeline \cite{brooks19unprocessing} to synthesize 
an unlimited number of blurry and noisy raw images.

\section{Image formation model}

\subsection{Camera pipeline}
The overall image acquisition pipeline
is summarized in Figure~\ref{fig:forward_model}.
On our model, we start from an idealized, sharp and
wavelength-dependent irradiance function $X$ defined over the continuous
image domain. The optics of the camera transform it into a blurry
function $B$, which is then digitized into a raw image $y$ with spatial,
temporal and spectral integration processes corresponding to the pixel
extent, the exposure time (including motion blur), and the Bayer pattern
typically used in digital cameras.The raw image is then a {\em
demosaicked} linRGB image $d$ interpolating the missing color channels
(see, for example, \cite{li08survey}), before being converted through the ISP
pipeline into and sRGB image $b^\prime$. It is impossible to recover the
continuous-domain function $X$ from discrete measurement. We thus
estimate instead a sharp digital $x$ mapping onto $y$ through an
approximation of the actual image formation process.

 \subsection{Approximate forward model}

In Figure \ref{fig:forward_model}, the matrices $K$ and $M$ approximate
respectively the blur caused by the optics, motion and defocus for the former,
and the sampling of colors in the RGB blurry image by the sensor in the latter.
The approximate forward model represented in the yellow box of 
Figure \ref{fig:forward_model} thus reads
\begin{equation}\label{eq:formationmodel}
    y = MKx + \varepsilon \quad \text{with} \quad \varepsilon \sim 
    \mathcal{N}(0, \lambda_{\text{s}} MKx + \lambda_{\text{r}}),
\end{equation}
where, according to \cite{foi08practical, brooks19unprocessing},
the vector $\varepsilon$ is the sensor's noise seen on a raw or linRGB 
image can be modelled as a pixel-varying zero-mean Gaussian distribution 
with variance $\lambda_{\text{s}} MKx + \lambda_{\text{r}}$.
It linearly depends on $MKx$ and two scalars: 
a scaling weight $\lambda_\text{s}$ and an
offset $\lambda_\text{r}$ representing respectively
the impact of shot and read noise \cite{foi08practical}.

This model contrasts with classical approaches for non-blind deblurring
\cite{chen17trainable, kruse17learning, zoran11from, eboli20end2end, zhang20deep}
simply considering a blur matrix $K$
between the sharp and blurry sRGB images $x^\prime$ and $b^\prime$ in the red 
box of Figure \ref{fig:forward_model}, stored on a device.
 
The traditional forward model for image deblurring assumes that the known blurry sRGB image $b^\prime$
has been obtained by applying a linear operation, \eg the convolution with a linear blur kernel, to an unknown sharp sRGB image $x^\prime$.
However this approach is a simplification of the physical model causing blur on the
analog image $B$ and thus ignores the 
different, possibly non-linear, components and transformations in a camera and shown
in Figure \ref{fig:forward_model}.
A linear forward model linking blurry and sharp images
such as \eqref{eq:formationmodel} is thus only valid between linRGB images.
Nah \etal \cite{nah17deep} indeed restore linRGB blurry images 
but they assume to have access to the blurry image $b$, 
formed from the sharp image $x$ in Fig.~\ref{fig:forward_model}, 
but since we only have access to $y$, this approach should rather 
be applied on a demosaicked version $d$ of $y$
approximating $b$.

The raw image formation
model of Eq.~\eqref{eq:formationmodel} is
particularly well-suited for camera PSF removal since 
the properties of most recent lenses are 
accurately measured and tabulated\footnote{https://www.dxomark.com/Lenses/}
or can be estimated with calibration of a camera, \eg
\cite{schuler11correction, kee11modeling}, or with an
optimization-based technique, \eg \cite{yue15blind, schuler12blind}.
This ensures that $K$ is known for this task. The mosaicking pattern
in $M$ is a feature of the camera and can reasonably be supposed known too in general.

\section{Proposed approach}

A natural approach for solving a joint deblurring, demosaicking and denoising problem
is to leverage the important previous work
on image demosaicking and denoising 
and non-blind RGB image deblurring 
by using a two-stage method stacking
a joint demosaicking and denoising solver followed by a non-blind deblurring approach,
\eg \cite{schuler11correction}, as shown in Figure \ref{fig:forward_model}.
One of the main contributions of this work 
is to instead predict a sharp, demosaicked and denoised image from the 
observation $y$.

\subsection{Energy function and splitting strategies}

We integrate 
the forward model \eqref{eq:formationmodel} into a penalized energy function
with an image prior $\Omega$ whose solution is a denoised and deblurred linRGB image:
\begin{equation}\label{eq:problemjoint}
    \min_x   \frac{1}{2}\left\| y - MKx\right\|_F^2 + \lambda \Omega(x).
\end{equation}
Optimization of \eqref{eq:problemjoint} is traditionally carried out with
splitting algorithms such as half-quadratic splitting (or HQS)~\cite{geman92constrained}.
We introduce an auxiliary
variable $z$ and solve
\begin{equation}
    \min_{x,z}   \frac{1}{2}\| y - MKz \|_F^2 + \lambda \Omega(x) \quad s.t. \quad z = x,
\end{equation}
which becomes, when relaxed with a weight $\beta >0$:
\begin{equation}\label{eq:problemjointdeblurring}
    \min_{x,z}   \frac{1}{2}\| y - MKz \|_F^2
        + \frac{\beta}{2} \| z - x \|_F^2 + \lambda \Omega( x ).
\end{equation}
Optimization requires to jointly handle the operators $M$ and $K$.
We will detail the calculations in the next paragraphs.

Alternatively, we first
demosaick the image, for instance with a CNN $\xi$ with parameter $\nu$ \cite{gharbi16deep},
and second use a non-blind deblurring 
approach on the demosaicked linRGB image to predict the final sharp linRGB image $x$.
The same relaxation of a constraint on an auxiliary variable $z$ leads to
\begin{align}\label{eq:problemdirectdeblurring}
    & \min_{x, z} \frac{1}{2}\| d - Kz \|_F^2
        + \frac{\beta}{2} \| z - x \|_F^2 + \lambda \Omega( x ),\\
          & \text{with}\quad d  = \xi_{\nu}(y, \lambda_\text{r}, \lambda_\text{s}).\nonumber
\end{align}
The demosaicking approach takes as input the noise parameters $\lambda_\text{r}$ and
$\lambda_\text{s}$, in the vein of \cite{gharbi16deep, kokkinos19iterative}.
In this case, the reference image in the non-blind deblurring problem is $d$ and not $y$.

\subsection{Solving the intermediate problems}

Predicting $x$ in both Eqs~\eqref{eq:problemjointdeblurring} and \eqref{eq:problemdirectdeblurring} is done by solving
\begin{equation}\label{eq:proximalstep}
    \min_{x} \lambda \Omega(x) + \frac{\beta}{2}\| z - x \|_F^2.
\end{equation}
The minimizer of energy can be computed by
evaluating in $z$ the proximal operator $\phi$ of $\Omega$  with
parameter $\lambda / \beta$ \cite{parikh14proximal}: 
\begin{equation}
    x = \text{prox}_\Omega(z,\lambda / \beta) = \phi(z, \lambda/\beta).
\end{equation}
The intermediate deblurred image in~\eqref{eq:problemdirectdeblurring} is the solution of:
\begin{equation}\label{eq:nonblinddeblurring}
    \min_z  \| d - Kz \|_F^2 + \beta \| z - x \|_F^2,
\end{equation}
which is classically solved with fast Fourier transform (FFT) \cite{zhang20deep}, 
conjugate gradient (CG) \cite{pulli14flexisp},
or Richardson fixed point iterations \cite{eboli20end2end}.

However estimating $z$ in Eq.~\eqref{eq:problemjointdeblurring} requires 
solving instead:
\begin{equation}\label{eq:demoisaicking}
    \min_z  \| y - MKz \|_F^2 + \beta \| z - x \|_F^2,
\end{equation}

For any mosaicking pattern choice, 
$M$ independently samples the color channels resulting in
\begin{equation}\label{eq:MKoperator}
    MKz = 
    \begin{bmatrix}
    D_RK_R & 0 & 0\\
    0 & D_GK_G & 0\\
    0 & 0 & D_BK_B\\
    \end{bmatrix} 
    \begin{bmatrix}
    z_R\\
    z_G\\
    z_B\\
    \end{bmatrix} 
\end{equation}
where $z_R$, $z_G$ and $z_B$ are the red, green and blue components and
$K_R$, $K_G$ and $K_B$ are the color-specific components of the blur
$K$.
One of our contributions is to estimate $z_R^\star$, $z_G^\star$ and $z_B^\star$,
the red, green and blue images forming $z^\star$, the solution of ~\eqref{eq:demoisaicking}, 
with a FFT-based module similar to the one of \cite{zhang20deep} in the case of the Bayer pattern.

\subsection{FFT-based solver for least-squares~\eqref{eq:demoisaicking}}
\label{sec:fft}

\begin{figure}[t]
    \centering
    \includegraphics[scale=0.5]{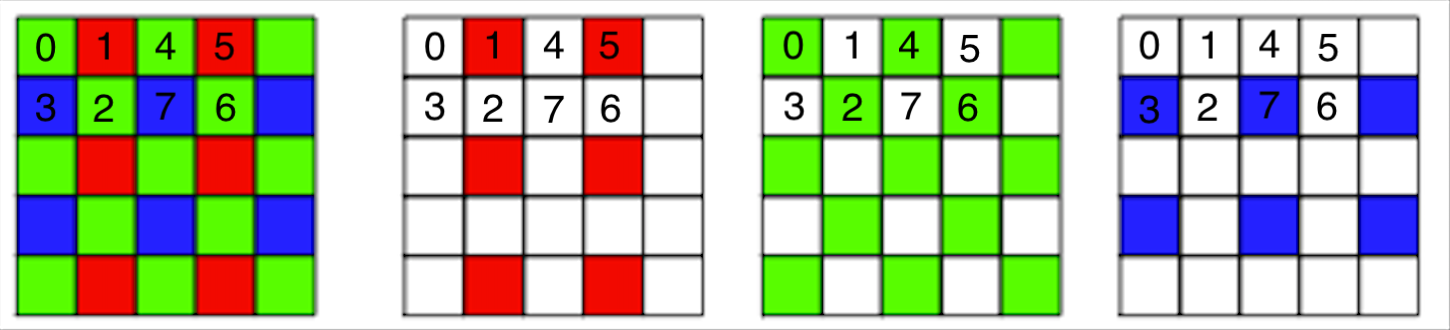}
    \caption{A mosaicked image with the Bayer pattern and its 
    three sampled colored components.}
    \label{fig:bayer}

\end{figure}

We solve Eq.~\eqref{eq:demoisaicking} with a FFT-based module inspired 
by \cite{zhang20deep} in the context of image upsampling.
The linear operator $MK$ in \eqref{eq:MKoperator} 
decomposes the least-squares problem Eq.~\eqref{eq:demoisaicking} into 
three independent terms.
Figure \ref{fig:bayer} shows a mosaicked image $y$ obtained sampled with $M$ and whose $y_R$ and $y_B$ components
are sampled versions of the corresponding RGB image
where in each $2\times2$ non-overlapping patch, only one
pixel value is retained per color.
Similarly, the $y_G$ is the sum of two
images $y_{G_1}$ and $y_{G_2}$ 
such that $y_G = y_{G_1} + y_{G_2}$ 
(and $D_G = D_{G_1} + D_{G_2}$), each one sampling a single
green pixel in the $2\times2$ non-overlapping patches
in Fig.~\ref{fig:bayer}.
We interpolate the missing values by solving,
with $c$ in $\{G_1,R,G_2,B\}$:
\begin{equation}\label{eq:demoisaickingcolor}
    \min_{z_c} \|y_c - D_c K_c z_c \|_F^2 + \beta \| z_c - x_c \|_F^2.
\end{equation}
These four problems are similar to the upsampling approach
of \cite{zhang20deep} with rate 2 solving the 
linear system, with $c$ in $\{G_1,R,G_2,B\}$:
\begin{equation}\label{eq:ls}
    (K_c^\top D_c^\top D_c K_c + \beta I )z_c^\star = K_c^\top D_c^\top y + \beta x_c.
\end{equation}
By splitting the green image into $y_{G_1}$ and $y_{G_2}$, 
we efficiently solve the four least-squares 
with an adapted version
of the FFT-based approach of \cite{zhao16fast, zhang20deep}.
We detail the implementation details and modifications on 
the code of \cite{zhang20deep} in the supplemental material.
The images $z_R^\star$ and $z_B^\star$ are the solutions of \eqref{eq:ls}
and $z^\star_G$ is obtained from $z_{G_1}^\star$ and $z_{G_2}^\star$ as 
follows: 
the pixels at locations $0,4,\dots$ 
(resp. $2,6,\dots$) in Fig.~\ref{fig:bayer}
are copied from the ones from $z^\star_{G_1}$ 
(resp. $z^\star_{G_2}$) and
the remaining pixels at locations $1,3,5,\dots$
are the corresponding values in $(z^\star_{G_1} + z^\star_{G_2})/2$.
Stacking the three images $z^\star_R$, $z^\star_G$ and $z^\star_B$
yields the RGB solution $z^\star$ of Eq.~\eqref{eq:demoisaicking}.

\subsection{Learnable embedding}

We improve the performance of the restoration method for solving either \eqref{eq:problemdirectdeblurring} or \eqref{eq:problemjointdeblurring},
we embed a few stages of HQS in the USRNet model of Zhang \etal \cite{zhang20deep} featuring two
modules for learning the proximal step \eqref{eq:proximalstep} and 
estimating on-the-fly the optimal weights $\beta^{(t)}$ 
and $\gamma^{(t)} = \lambda/\beta^{(t)}$ $(t=1,\dots,T)$.
We parameterize the proximal 
operator of $\phi$ with the same Unet
model as in \cite{zhang20deep}
with parameter $\theta$ such that for a given estimate $z^{(t)}$,
we predict an RGB image $x$ as
\begin{equation}\label{eq:phifunction}
    x^{(t+1)} = \phi_\theta(z^{(t)}, \gamma^{(t+1)}).
\end{equation}
We predict $z^{(t+1)}$ from $x^{(t+1)}$, $K$, $M$ and $\beta^{(t+1)}$ with the mapping $\psi$ estimating its $R$, $G$ and $B$ components 
with the approach of Section \ref{sec:fft}:
\begin{equation}\label{eq:psifunction}
    z^{(t+1)} = \psi(x^{(t+1)}, K, M, \beta^{(t+1)}),
\end{equation}
where $x^{(t+1)}$, $K$, $M$ and $\beta^{(t+1)}$ are used to build the
least-squares problems \eqref{eq:ls}.
The weights
$\beta^{(t)}$ and $\gamma^{(t)}$ with a 3-layer perceptron,
as detailed in \cite{zhang20deep} dubbed $\chi$ 
and with parameter $\omega$.
In our case, it becomes a function of the read and shot noise
coefficients $\lambda_\text{r}$ and $\lambda_\text{s}$ defined as:
\begin{equation}
    [ \{\beta^{(t)}\}_{t=1}^T, \{\gamma^{(t)}\}_{t=1}^T] = \chi_\omega(\lambda_\text{r},
    \lambda_\text{s}).
\end{equation}

\begin{figure}
    \begin{minipage}[t]{0.48\textwidth}
    \centering
    \begin{algorithm}[H]
    \SetAlgoLined
    \KwData{$y$, $K$, $M$, $\lambda_\text{r}$, $\lambda_\text{s}$, $\theta$, $\omega$}
        \tcp{Weights prediction}
     $[ \{\beta^{(t)}\}_{t=1}^T, \{\gamma^{(t)}\}_{t=1}^T] = \chi_\omega(\lambda_\text{r},
        \lambda_\text{s})$\;
        \tcp{Demosaicking}
        $d\leftarrow\xi_\nu(y,\lambda_\text{r},
        \lambda_\text{s})$\;
        $t \leftarrow 1$\;
        $x \leftarrow d$\;
        \tcp{Restoration with variant of \cite{zhang20deep}}
     \For{$t \leq T$}{
      $z \leftarrow \psi(y, x, K, M, \beta^{(t)})$\;
      $x \leftarrow \phi_\theta(z, \gamma^{(t)})$\;
      $t \leftarrow t+1$\;
     }
     \KwResult{Restored linRGB image $x$.}
     \caption{Parametric function for solving \eqref{eq:problemjointdeblurring}.}
     \label{alg:usrnetjoint}
    \end{algorithm}
    \end{minipage}%
    \hfill
    \begin{minipage}[t]{0.48\textwidth}
    \centering
    \begin{algorithm}[H]
    \SetAlgoLined
    \KwData{$y$, $K$, $\lambda_\text{r}$, $\lambda_\text{s}$, $\theta$, $\omega$, $\nu$}
        \tcp{Weights prediction}
     $[ \{\beta^{(t)}\}_{t=1}^T, \{\gamma^{(t)}\}_{t=1}^T] = \chi_\omega(\lambda_\text{r},
        \lambda_\text{s})$\;
        \tcp{Demosaicking}
        $d\leftarrow \xi_\nu(y,\lambda_\text{r},
        \lambda_\text{s})$\;
        $t \leftarrow 1$\;
        $x \leftarrow d$\;
        \tcp{Deblurring with \cite{zhang20deep}}
     \For{$t \leq T$}{
      $z \leftarrow \psi(d, x, K, I, \beta^{(t)})$\;
      $x \leftarrow \phi_\theta(z, \gamma^{(t)})$\;
      $t \leftarrow t+1$\;
     }
     \KwResult{Restored s/linRGB image $x$.}
     \caption{Parametric function for solving \eqref{eq:problemdirectdeblurring}.}
     \label{alg:usrnetdirect}
    \end{algorithm}
    \end{minipage}%
\end{figure}

Our proposed approach for joint deblurring, demosaicking and denoising of raw images
embeds the weight predictor $\chi_\omega$,
the learnable proximal operator $\phi_\theta$ and the FFT-based
solver for joint deblurring and demosaicking $\psi$ 
in \eqref{eq:psifunction}
into the state-of-the-art model USRNet \cite{zhang20deep} 
initially designed for sRGB image non-blind deblurring and upsampling.

The two-stage approach first jointly demosaicks and denoises a raw image with
a module $\xi_\nu$ that we implement with the learning-free approach of \cite{menon07demosaicing} or the state-of-the-art approach of \cite{gharbi16deep} dubbed Deepjoint.
It is followed by a non-blind deblurring module that we implement with the 
USRNet model of \cite{zhang20deep}.
We modify Deepjoint and USRNet for processing the noise parameters $\lambda_\text{s}$ and 
$\lambda_\text{r}$ in place of the variance of a traditional Gaussian noise model.
These two approaches are summarized in
Algorithms~\ref{alg:usrnetjoint} and \ref{alg:usrnetdirect}.
The main difference is that in Alg.~\ref{alg:usrnetjoint}, our approach
takes $y$ as reference during inference whereas Alg.~\ref{alg:usrnetdirect}
uses $y$ only in the joint demosaicking and denoising module.
In this case, the deblurring module could predict wrong details as it
takes the demosaicked image $d$ which might contain
prediction errors such as Moir\'e artifacts.

\section{Experiments}

We run the experiments on an NVIDIA Tesla V100 graphic card.
The code will be made available if the paper is accepted.

\paragraph{Experimental setting.}
We extract $96\times96$ patches in the training images of DIV2K and Flickr2K
datasets, often used for training image upsampling models and
featuring high-resolution
edges, Moir\'e artifacts-prone textures and little compression artifacts.
We convert these patches into the linRGB format
with the pipeline of \cite{brooks19unprocessing}, blur
them we with uniform blur from \cite{zhang20deep}
and add affine noise
generated with the code of \cite{brooks19unprocessing}.
We randomly flip and rotate the patches as augmentation.
We extract 5000 patches from the 100 images of DIV2K validation set to form ours.
Since blur is color-dependent \cite{schuler11correction, schuler12blind, yue15blind},
we synthesize RGB blur kernels (details in what follows) that are more realistic
than grayscale filters such as the ones of \cite{levin09understanding}.
We use Adam optimizer with initial learning rate set to $10^{-4}$.
The learning rate is divided by 2 whenever the 
validation loss plateaus during 15 epochs until reaching $10^{-6}$.
We use a batch size of 32.
We use the $\ell_1$ loss to compare the ground-truth patches and the predictions in the sRGB format as done in \cite{brooks19unprocessing}.

\paragraph{Generating color-specific blur kernels.}
Real blurs are color-specific because of the diffraction
effect of a lens. This can be seen on the PSFs estimated
by \cite{schuler11correction}, which locally is modelled
with an RGB uniform kernel whose components are
smoothly varying across color channels \cite{schuler11correction, yue15blind}.
We thus build RGB kernels 
for training, validating and testing the models in a more 
realistic setting than traditional Levin's kernels \cite{levin09understanding} 
for instance.
We generate a $25 \times 25$ grayscale kernels with the code 
of Zhang \etal \cite{zhang20deep} representing both motion
and anisotropic Gaussian blurs.
This synthetic kernel is arbitrarily set as the blue component.
The other ones are randomly rotated versions with small angles in
$[-5^\circ, +5^\circ]$ and rescaled by a factor~$[0.8,1]$.
This yields a $25 \times 25 \times 3$ array.

\begin{table*}[t]
    \adjustbox{max width=\textwidth}{%
    \centering
    \begin{tabular}{lcccccccc} \toprule
        \multicolumn{1}{l}{Datasets}
         & \multicolumn{4}{c}{Kodak (192 images)} & \multicolumn{4}{c}{Sun (640 images)} \\ 
        \multicolumn{1}{l}{Noise level}
         & \multicolumn{2}{c}{Noiseless} & \multicolumn{2}{c}{With noise}  & \multicolumn{2}{c}{Noiseless} & \multicolumn{2}{c}{With noise}\\ 
        \multicolumn{1}{l}{Color space}
         & linRGB & sRGB & linRGB & sRGB  & linRGB & sRGB & linRGB & sRGB\\ \midrule
        \cite{menon07demosaicing} + \cite{krishnan09fast} & 
        33.15/0.91 & 
        28.03/0.78 & 
        30.94/0.79 & 
        22.77/0.47 & 
        36.55/0.93 & 
        29.99/0.82 & 
        32.57/0.79 & 
        23.59/0.50\\
        \cite{menon07demosaicing} + \cite{zhang20deep} (s/s) & -~/~- & 
        32.16/0.88 & 
        -~/~- & 
        29.56/0.78 & 
        -~/~- & 
        33.50/0.91 & 
        -~/~- & 
        30.40/0.80\\
        \cite{menon07demosaicing} + \cite{zhang20deep} (lin/lin) & 
        \underline{35.92}/\underline{0.94} & 
        31.92/0.89 & 
        33.74/\underline{0.90} & 
        29.42/0.78 & 
        \underline{39.43}/\underline{0.96} & 
        33.77/0.91 & 
        36.27/0.92 & 
        30.46/0.81 \\
        \cite{menon07demosaicing} + \cite{zhang20deep} (lin/s) & 
        35.28/\textbf{0.95} & 
        32.28/0.90 & 
        33.28/\underline{0.90} & 
        29.69/\underline{0.79} & 
        38.73/\underline{0.96} & 
        34.21/0.92 & 
        35.83/\textbf{0.92} & 
        30.66/\underline{0.81}\\
        \cite{gharbi16deep} + \cite{krishnan09fast}& 
        32.85/0.90 & 
        27.89/0.75 & 
        32.10/0.87 & 
        26.87/0.69 & 
        36.07/0.92 & 
        29.69/0.80 & 
        34.92/0.90 & 
        28.26/0.73\\
        \cite{gharbi16deep} + \cite{zhang20deep} (s/s) & 
        -~/~- & 
        30.27/0.83 & 
        -~/~- & 
        29.01/0.77 & 
        -~/~- & 
        31.73/0.86 & 
        -~/~- & 
        30.07/0.80\\
        \cite{gharbi16deep} + \cite{zhang20deep} (lin/lin) & 
        35.83/\underline{0.94} & 
        31.67/0.88 & 
        \underline{33.88}/\underline{0.90} & 
        29.48/\underline{0.79} & 
        38.97/\underline{0.96} & 
        33.24/0.90 & 
        \underline{36.11}/\textbf{0.92} & 
        30.35/\underline{0.81}\\
        \cite{gharbi16deep} + \cite{zhang20deep} (lin/s) & 
        35.04/\underline{0.94} & 
        32.12/0.88 &  
        33.12/\underline{0.90} & 
        29.61/0.78 & 
        37.98/\underline{0.96} & 
        33.61/0.90 & 
        35.54/\underline{0.91} & 
        30.58/\underline{0.81} \\
        \hdashline
        Ours (lin/lin)  & 
        \textbf{36.48}/\textbf{0.95} & 
        32.46/0.90 & 
        \textbf{34.10}/\textbf{0.91} & 
        29.76/\textbf{0.80} &  
        \textbf{40.10}/\textbf{0.97} & 
        34.39/0.93 & 
        \textbf{36.51}/\textbf{0.92} & 
        30.72/\textbf{0.82} \\
        
        Ours (lin/s) & 
        35.72/\textbf{0.95} & 
    \underline{32.99}/\underline{0.91} & 
    33.52/\textbf{0.91} & 
    \underline{29.98}/\textbf{0.80} & 
    39.13/\textbf{0.97} & 
    \underline{34.90}/\underline{0.93} & 
    35.93/\textbf{0.92} & 
    \underline{30.86}/\textbf{0.82}\\
        Ours (lin/s, gray) & 
        35.08/\underline{0.94} & 
        32.21/0.89 & 
        33.26/\underline{0.90} & 
        29.72/\underline{0.79} & 
        38.48/\underline{0.96} & 
        34.15/0.92 & 
        35.81/\textbf{0.92} & 
        30.71/\textbf{0.82} \\
        Ours (lin/s,  \cite{gharbi16deep})  & 
        35.86/\textbf{0.95} & 
        \textbf{33.37}/\textbf{0.92} & 
        33.53/\textbf{0.91} & 
        \textbf{30.14}/\textbf{0.80} & 
        39.21/\textbf{0.97} & 
        \textbf{35.19}/\textbf{0.94} & 
        35.87/\textbf{0.92} & 
        \textbf{30.95}/\textbf{0.82}\\
        \bottomrule
    \end{tabular}}
    \caption{Joint deblurring, denoising and demosaicking comparison. Best result is in bold font. Second best is underlined.}
    \label{tab:deblurring}
\end{table*}

\begin{figure*}
    \centering
    \adjustbox{max width=0.99\textwidth}{
    \begin{tabular}{ccc}
        \begin{subfigure}[b]{0.33\textwidth}
         \centering
         \includegraphics[width=\textwidth]{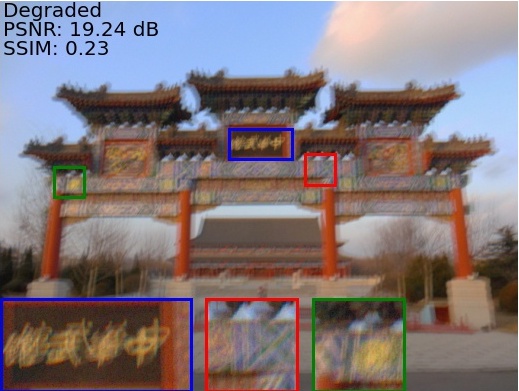}
     \end{subfigure}
         & 
         \begin{subfigure}[b]{0.33\textwidth}
         \centering
         \includegraphics[width=\textwidth]{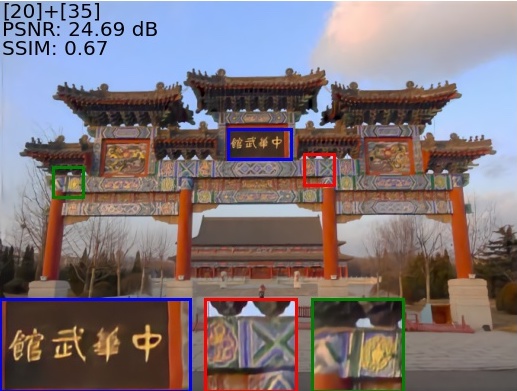}
     \end{subfigure}
        &
        \begin{subfigure}[b]{0.33\textwidth}
         \centering
         \includegraphics[width=\textwidth]{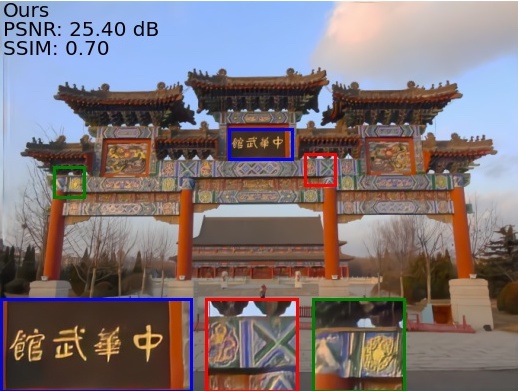}
         \end{subfigure}
        \\
        \begin{subfigure}[b]{0.33\textwidth}
         \centering
         \includegraphics[width=\textwidth]{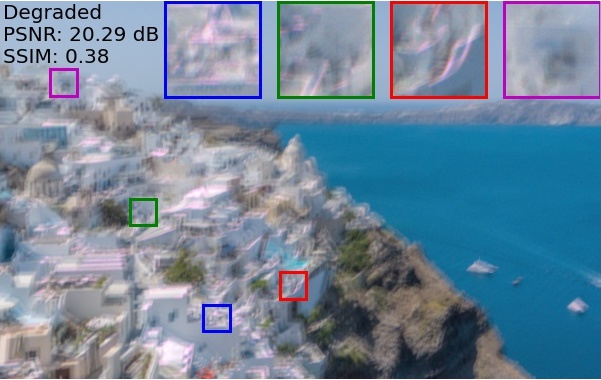}
     \end{subfigure}
         & 
         \begin{subfigure}[b]{0.33\textwidth}
         \centering
         \includegraphics[width=\textwidth]{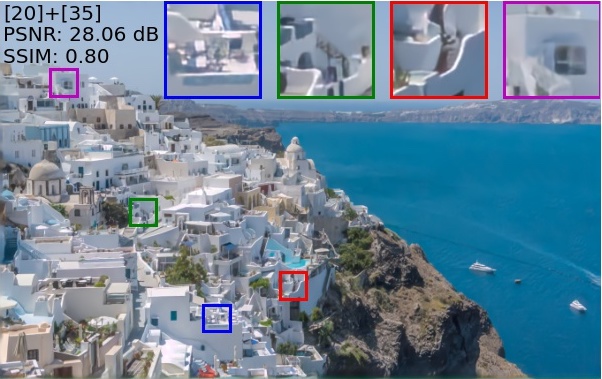}
     \end{subfigure}
        &
        \begin{subfigure}[b]{0.33\textwidth}
         \centering
         \includegraphics[width=\textwidth]{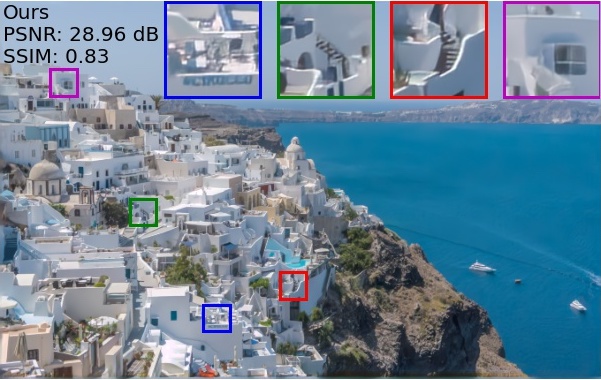}
     \end{subfigure}
    \end{tabular}
    }
    \caption{Examples of jointly deblurred, demosaicked and denoised images. We show the degraded raw images in the sRGB format.
    Compared to the two-stage method \cite{menon07demosaicing}+\cite{zhang20deep}, our method 
    restores finer details.}
    \label{fig:uniformdeblurring}
\end{figure*}

\begin{figure*}[h]
     \centering
     \begin{subfigure}[b]{0.24\textwidth}
         \centering
         \includegraphics[width=\textwidth]{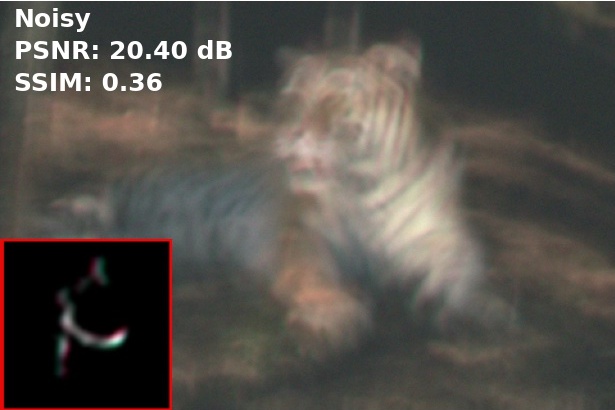}
     \end{subfigure}
     \hfill
     \begin{subfigure}[b]{0.24\textwidth}
         \centering
         \includegraphics[width=\textwidth]{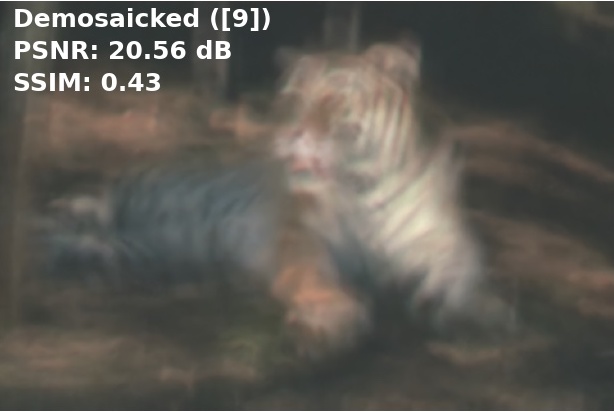}
     \end{subfigure}
     \hfill
     \begin{subfigure}[b]{0.24\textwidth}
         \centering
        \includegraphics[width=\textwidth]{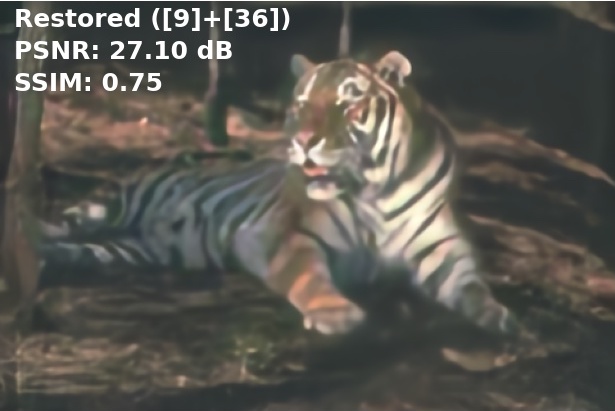}
     \end{subfigure}
     \hfill
     \begin{subfigure}[b]{0.24\textwidth}
         \centering
        \includegraphics[width=\textwidth]{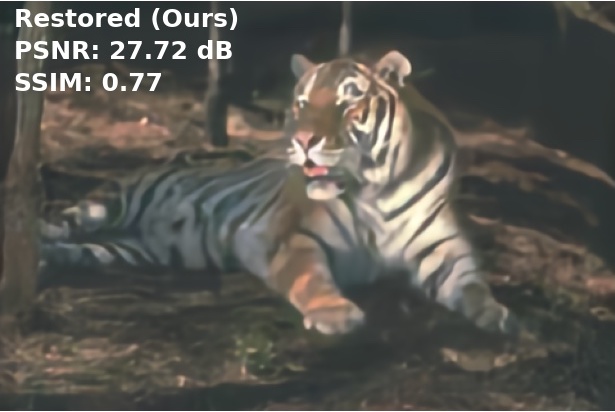}
     \end{subfigure}
        \caption{
        Restored exampled with a $65 \times 65$ blur kernel and  noise parameters set to $\lambda_\text{s}=10^{-3}$ and $\lambda_\text{r}=1.3\times10^{-6}$.
        Better seen on a computer screen. Both quantitatively and visually our method
        outperforms the two-stage method \cite{gharbi16deep}+\cite{zhang20deep}.}
        \label{fig:largerkernels}
\end{figure*}

\paragraph{Joint deblurring, denoising and demosaicking.}

We evaluate our method on two synthetic datasets. 
We convert the 24 and 80 images of the Kodak \cite{li08survey}
and Sun \cite{sun13edge} datasets into linRGB images 
with the pipeline of \cite{brooks19unprocessing} and
blur them with the 8 filters of \cite{levin09understanding}
we transform into RGB kernels.
We add noise with the code of \cite{brooks19unprocessing}
with $\log(\lambda_\text{s})$ chosen in [$10^4$, $3\times10^{-3}$] and
corresponding $\lambda_\text{r}$ in the model of \cite{brooks19unprocessing},
\ie small to moderate noise, and mosaick
them to form respectively 192 and 640 test samples.
We compare our approach on raw images with two-stage methods.
For demosaicking, we use either the filtering approach of \cite{menon07demosaicing}
or the CNN for joint demosaicking and denoising of \cite{gharbi16deep}.
We retrain \cite{gharbi16deep} to take into account the noise 
distribution of \cite{brooks19unprocessing} on the 2.5 million patches of \cite{gharbi16deep}.
For non-blind deblurring, we use \cite{krishnan09fast} based on a 
hyper-Laplacian image prior or the unrolled model of
\cite{zhang20deep}.
We retrain \cite{zhang20deep} each time on the images predicted by the first
stage implemented with \cite{gharbi16deep} or \cite{menon07demosaicing} to
take into account prediction errors.
As we predict linRGB images but ultimately want enhanced sRGB images, with
compare three kinds of supervision: (i) the intermediate image $d$ is converted
into an sRGB image and \cite{zhang20deep} is supervised with sRGB targets (s/s), 
(ii) \cite{zhang20deep} deblurs a linRGB demosaicked image and is supervised with linRGB targets
(lin/lin), and (iii) \cite{zhang20deep} deblurs a linRGB demosaicked image and 
is supervised with sRGB targets as in \cite{brooks19unprocessing} (lin/s).
We train a variant of our model in the ``lin/lin'' setting and three in the ``lin/s''
setting: one with initial guess demosaicked with simple bilinear interpolation, 
one with initial guess obtained with \cite{gharbi16deep} (lin,s \cite{gharbi16deep})
and one trained with grayscale 
kernels only (lin/s, gray).
We unroll $T=6$ iterations of HQS in the different implementations of \cite{zhang20deep}
in 
Algorithms \eqref{alg:usrnetjoint} and \eqref{alg:usrnetdirect} implemented by
the baselines and our approach.

Table \ref{tab:deblurring} shows the PSNR and SSIM scores on the images after we 
crop 50 pixels on the borders to discard any boundary artifact in the measurements.
Our methods achieve on the four sets the best PSNR/SSIM score by PSNR margins of 0.5dB and 
SSIM margins of 0.01 or 0.02 over the two-stage methods.
Methods trained with supervision on linRGB images naturally have the best scores on this color
format but are behind the other methods in the sRGB format, suggesting supervision
with sRGB sharp images 
with the approach of \cite{brooks19unprocessing} is also beneficial for deblurring and 
demoisacking raw degraded images.
Our variant trained only on grayscale kernels is in the ballpark of the best ones in the noisy 
cases but lags behind them by margins of 1dB in the noiseless case, meaning it cannot restore
as fine details as the methods trained with the same blur distribution.
The table also shows that initialization matters as the method with initial guess produced 
by \cite{gharbi16deep} leads to better results, with margins ranging from 0.1 to 0.4dB 
on sRGB images, compared to the
one initialized with a demosaicked image bilinearly interpolated.
Figure \ref{fig:uniformdeblurring} shows two restoration examples of 
blurry, mosaicked and noisy raw images (displayed as sRGB images) obtained
with our best performing method and the best performer from the two-stage
techniques.
We also provide comparison with the same images but with the classical grayscale kernels of
\cite{levin09understanding} in the supplemental material. 
The method is as fast as vanilla USRNet \cite{zhang20deep} since the only
modification that might alter running time is the FFT-based module
whose computation time is negligible compared to evaluating a CNN.
It takes about 1 second to process a 720p image and at most 5 seconds
for a 2K image.

\paragraph{Robustness to larger kernels.}
We train the different models with $25 \times 25$ kernels but we show in
Fig.~\ref{fig:largerkernels} that our method can be used with much larger filters.
The image is blurred with a $65\times65$ kernel from \cite{pan18dark}. 
We compare the two-stage strategy \cite{gharbi16deep}+\cite{zhang20deep} 
to ours, both trained with the ``lin/lin'' setting. Our method achieves 
a better PSNR score and visual aspect compared to the two-stage method.
This is typical of our observations on other large kernels from \cite{pan18dark}.

\begin{figure*}
    \centering
    \begin{tabular}{cc}
        \begin{subfigure}[]{0.48\textwidth}
             \centering
             \includegraphics[width=\textwidth]{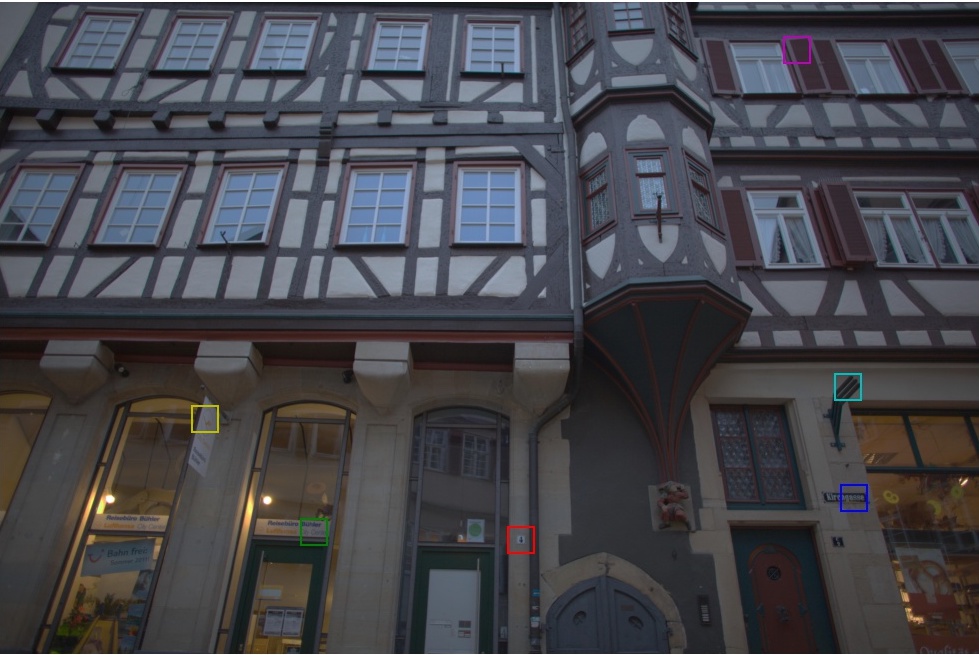}
         \end{subfigure} &
         \begin{subfigure}[]{0.48\textwidth}
             \centering
             \includegraphics[width=\textwidth]{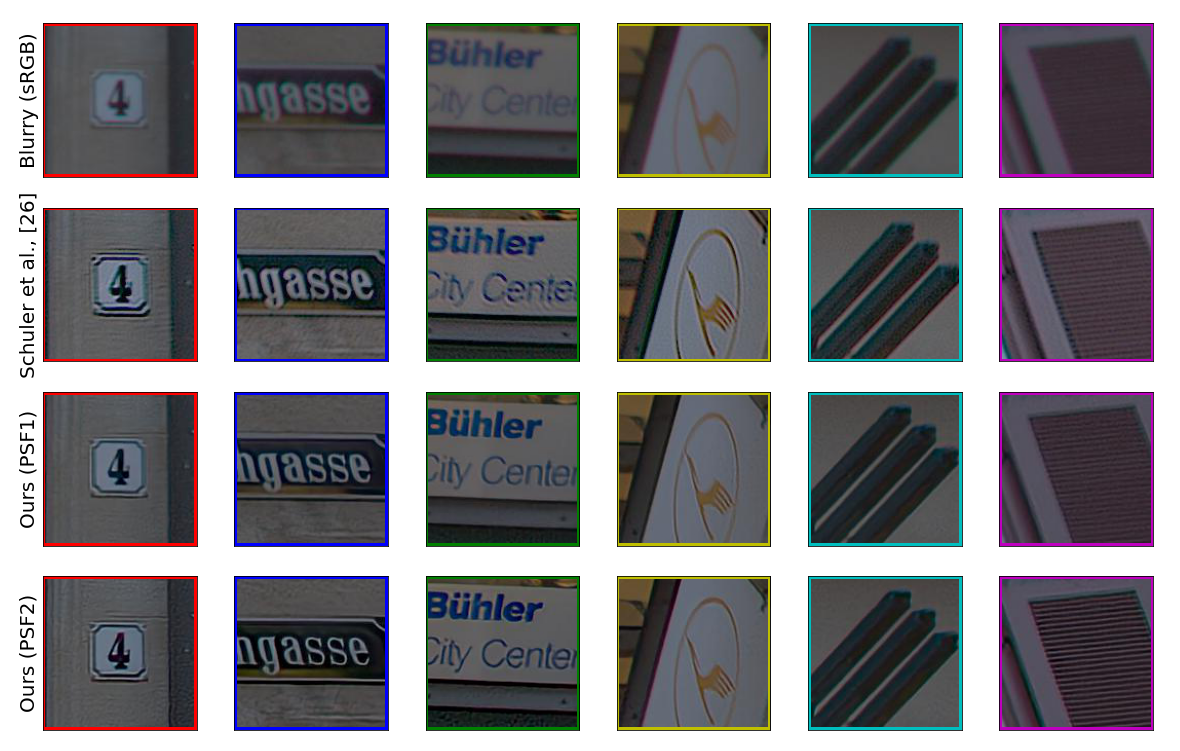}
         \end{subfigure}
    \end{tabular}
    \caption{The real blurry image of \cite{schuler11correction} 
    was taken with a modern digital SLR and a zoom lens at
    maximal aperture, exhibiting chromatic aberrations,
    especially on the corners where the lens is the most curved.
    Our method 
    efficiently removes
    the blur caused by the optics, provided either a calibrated or an approximate PSF (best seen on a computer screen).}
    \label{fig:facadepsf}
\end{figure*}

\paragraph{Camera PSF removal.}
We now remove the aberrations of a consumer-grade lens PSF.
We restore the blurry image\footnote{https://www.webdav.tuebingen.mpg.de/pixel/lenscorrection/}
shot with a Canon Mark II reflex camera and a canon 24mm f/1.4 lens at maximal aperture and whose
PSF has been measured by two approaches\footnote{https://www.webdav.tuebingen.mpg.de/pixel/blind\_lenscorrection/}: a calibration method \cite{schuler11correction} 
and a variational approach \cite{schuler12blind}.

We convert the corresponding sRGB real blurry image into a raw image with 
the camera pipeline of \cite{brooks19unprocessing}.
We follow \cite{schuler11correction, schuler12blind} and
 break the full image into overlapping patches
where the PSF boils down to a locally uniform blur kernel.
We restore each patch with our model trained for the previous experiment
without fine-tuning it with the PSF, stitch them together 
as detailed in \cite{schuler11correction} and convert 
the restored image back into the sRGB format.
We show in Figure \ref{fig:facadepsf} the results for the PSF obtained with
camera calibration (PSF1) and the one predicted with a variational method
(PSF2). We compare them to the image restored in \cite{schuler11correction}
that also removes blur from the raw image provided with the PSFs.
Our methods can restore finer details such as the words on the panels
or the closest in
Figure \ref{fig:facadepsf}, with both PSFs.
We provide other examples of PSF removal 
from real images shot with the same lens 
in the supplemental material.

We also produce quantitative results on synthetic data 
in Table \ref{tab:psfdeblurring}. Since there is no existing pairs of blurry and sharp
images from modern SLRs, we collect the 53 images with ratio height/width of 2/3 
from the DIV2K validation dataset matching the ratio of the PSFs measured
by \cite{schuler11correction, schuler12blind}.
We convert them into linRGB images with \cite{brooks19unprocessing}, 
blur them with the PSF obtained by calibration of the camera
in \cite{schuler11correction} 
(which thus becomes the {\em ground-truth} blur for the synthetic data),
mosaick them and finally add affine noise with the same parameters as in previous experiment.
We evaluate the methods from Tab.~\ref{tab:deblurring} trained in the (lin/s) 
setting {\em without} further retraining them on the PSFs of \cite{schuler11correction, schuler12blind}.
We restore the images with the PSF from \cite{schuler11correction}
used to generate the blurry images 
and dubbed ``GT'' in Tab.~\ref{tab:psfdeblurring} 
and the one from \cite{schuler12blind}, considered as 
an approximate blur we call ``Approx.'' in Tab.~\ref{tab:psfdeblurring}.

\begin{table}[]
    \centering
    \begin{tabular}{lcccc} \toprule
        PSFs & \multicolumn{2}{c}{GT \cite{schuler11correction}} & \multicolumn{2}{c}{Approx. \cite{schuler12blind}} \\
         & linRGB & sRGB & linRGB & sRGB \\ \midrule
         \cite{menon07demosaicing}+\cite{zhang20deep} & 41.22 & 35.22 & \textbf{36.20} & \textbf{29.87} \\
         \cite{gharbi16deep}+\cite{zhang20deep} & 40.86 & 35.04 & 35.72 & 29.50\\
         \hdashline
         Ours & 41.14 & \underline{35.77} & 35.53 & 29.28 \\
         Ours (gray) & \textbf{41.40} & 35.63 & \underline{36.13} & \underline{29.76} \\
        Ours (\cite{gharbi16deep})& \underline{41.34} & \textbf{35.84} & 35.47 & 29.21 \\
        \bottomrule
    \end{tabular}
    \caption{PSF removal comparison.}
    \label{tab:psfdeblurring}
\end{table}

We achieve the best PSNR score with the ``GT'' PSF used to build the synthetic images
with margins of +0.6dB on the two-stage methods
and a margin of 0.2dB over the variant 
trained uniquely on grayscale kernels. 
However the two-stage techniques achieve better results
with the ``Approx.'' PSF by a margin of 0.6dB over our
methods trained on RGB kernels and a small margin of less than 
0.1dB over our method trained on grayscale kernels.
It suggests that our synthetic RGB kernels help to improve the 
performance of our joint restoration model when the PSF is 
accurately known.
This is a reasonable assumption 
provided professional lens benchmarks.
However the drop of performance with approximate kernels
could be explained by our model overfitting colored
filters that might not have a realistic distribution
permitting robustness to large prediction errors in the blur.
On the one hand, we could be able to improve 
results on approximate blurs in Tab.~\ref{tab:psfdeblurring}
with more realistic models of RGB kernels but
on the other hand, the example in Figure~\ref{fig:facadepsf} 
shows that our method can already handle predicted blurs in 
real-world scenarios.

\section{Conclusion}

We have presented a approximate forward model for joint deblurring,
demosaicking and denoising images, derived from a digital
camera pipeline.
We have proposed a penalized energy based on it, solved with HQS.
Iterations of this method are embedded into a parametric function
inspired by \cite{zhang20deep},
restoring raw images and supervised with the technique of \cite{brooks19unprocessing}.
Our experiments have shown that it outperforms two-stage approaches,
decomposing the problem into a demosaicking step followed 
by non-blind deblurring, quantitatively and visually
and in the presence of affine noise.
Our approach have been applied to the removal of
chromatic aberrations caused by the optics of 
a camera from real images, when the PSF is estimated beforehand.
Future work includes the generation of more realistic training data,
such as blur kernels, to remove more optical aberrations, \eg coma,
from raw images.

\paragraph{Acknowledgments:} This work was funded in part by the 
French government under management of Agence Nationale de la 
Recherche as part of the "Investissements d’avenir" program, 
reference ANR-19-P3IA-0001 (PRAIRIE 3IA Institute), the Louis 
Vuitton/ENS Chair in Artificial Intelligence and the Inria/NYU 
collaboration. 
Jian Sun was supported by NSFC (11971373, U20B2075, 12026605).

{\small
\bibliographystyle{ieee_fullname}
\bibliography{bib}
}

\end{document}